\documentclass[twoside,11pt]{article}

\def\degangle{\kern-.2em\r{}} 
\def\degC{\kern-.2em\r{}\kern-.3em C }
\usepackage[dvips]{graphicx}

\usepackage{cjp}

\cjpissue{unknown}{unknown}{unknown}
         {Unconventional superconductivity in Na$_{x}$CoO$_{2}\cdot y$H$_{2}$O}{H.\ Sakurai $et$ $al$.}{1}

\begin{document}


\title{Unconventional superconductivity in Na$_{x}$CoO$_{2}\cdot y$H$_{2}$O}

\author{Hiroya Sakurai$^{1}$, Kazunori Takada$^{2}$, Takayoshi Sasaki$^{2}$, and Eiji Takayama-Muromachi$^{1}$}

\address{
$^{1}$Superconducting Materials Center, National Institute for Materials Science (SMC/NIMS), 1-1 Namiki, Tsukuba, Ibaraki 305-0044, Japan.\\
$^{2}$Advanced Materials Laboratory, National Institute for Materials Science (AML/NIMS), 1-1 Namiki, Tsukuba, Ibaraki 305-0044, Japan/ CREST, Japan Science and Technology Agency (JST).
}
\date{unknown}
\maketitle              

\begin{abstract}
We synthesized powder samples of Na$_{x}$CoO$_{2}\cdot y$H$_{2}$O changing the volume of the water in the hydration process, then investigated their superconducting properties,. It was proved that the volume of water is one of key parameters to obtain a single phase of Na$_{x}$CoO$_{2}\cdot y$H$_{2}$O with good superconducting properties. The transition temperature, $T_{\mbox{c}}$, of the sample changed gradually while it was stored in the atmosphere of 70\% humidity. Superconducting behavior under high magnetic field was very sensitive to $T_{\mbox{c}}$. $H_{\mbox{c2}}$ of a high quality sample with high $T_{\mbox{c}}$ seemed very high.
\end{abstract}


\section{Introduction}

The cobalt oxide superconductor Na$_{x}$CoO$_{2}\cdot y$H$_{2}$O \cite{Nature} has characteristic features in its structure and Co valence state\cite{TakadaJMC}. The structure is composed of thick insulating layers of Na atoms and H$_{2}$O molecules, as well as conducting CoO$_{2}$ layers. In the CoO$_{2}$ layer, CoO$_{6}$ octahedra are connected to each other by edge sharing with the Co sites forming a triangular lattice. On the other hand, the valence of the Co ion is between 3+ (3$d^{6}$) and 4+ (3$d^{5}$) as in the case of Na$_{0.5}$CoO$_{2}$\cite{Singh}. Co$^{3+}$ and Co$^{4+}$ ions often have low spin states with S=0 and 1/2, respectively. At the early stage of research, this compound was considered as a system of S=1/2 triangular lattice doped with electrons and the doped resonating-valence-bond (RVB) model\cite{RVB} was expected to be valid for explaining its superconductivity. Now, it is widely recognized that the electron density of this compound is so large that the doped RVB model can not be applied simply. Most of studies reported thus far suggest strongly that the superconductivity of this compound is unconventional.

The symmetry of superconducting gap function is one of the most important parameters to elucidate the mechanism of superconductivity, and our understanding of superconductivity has become deeper with finding of various types of superconductivity with different symmetries.  Thus, many researchers have been greatly interested in the superconducting symmetry of Na$_{x}$CoO$_{2}\cdot y$H$_{2}$O since its discovery.

Theoretically, various possibilities have been proposed on the superconducting symmetry of the compound (or more generally, for the triangular lattice system), for example, $d$-, chiral $d$-, chiral $p$-, and $f$-waves\cite{sym}. These states can be distinguished from each other by microscopic experiments such as nuclear quadrupole resonance (NQR), muon spin rotation/relaxation ($\mu$SR), and nuclear magnetic resonance (NMR). In the NQR measurements, no Hebel-Schlichter peak (coherence peak) was observed in a $1/^{59}T_{1}T$-$T$ curve ($T_{1}$: relaxation time, and $T$: temperature)\cite{Ishida,Fujimoto}, which means that the superconducting gap has nodes on the Fermi surface. Thus, the symmetry of $s$-, chiral $d$-, and chiral $p$- waves are denied. It was pointed out previously for the case of chiral $d$- or chiral $p$-wave that if superconductivity is disturbed by impurities, the coherence peak is reduced and may not be observed\cite{ImpEffect}. However, an internal magnetic field should exist if chiral $d$ or chiral $p$ is the case, and it has never observed by $\mu$SR\cite{Higemoto}. These results are consistent with specific heat data, which suggest the presence of line-nodes\cite{Cp} in the superconducting gap. Therefore, the superconducting symmetry is limited to $d$- or $f$-wave. The $d$- and $f$-wave symmetries can be distinguished from $T$-dependence of Knight shift, $K$; sudden change of $K$ below $T_{\mbox{c}}$ is expected for spin singlet Cooper pairs with the $d$-wave symmetry, while invariable $K$ below $T_{\mbox{c}}$ along a certain direction for spin triplet pairing with the $f$-wave symmetry. Unfortunately, inconsistent results have been reported thus far on $K$ \cite{Kobayashi,Waki}. The NMR measurement usually needs an external magnetic field, and the magnetic field makes it difficult to obtain the real (intrinsic) Knight shift, that is, $K$ may apparently change below $T_{\mbox{c}}$ caused by demagnetization effect resulting from the superconducting diamagnetism, or on the contrary, $K$ may be apparently invariant if the applied magnetic field is close to or higher than the upper-critical field, $H_{\mbox{c2}}$. Thus, the estimation of $H_{\mbox{c2}}$ is quite important to interpret the NMR data.

Recently, many $H_{\mbox{c2}}$ data have been reported. According to magnetic measurements, we obtained an extremely high initial slope of $dH_{\mbox{c2}}/dT_{\mbox{c}}|_{H=0}=19.3$ T/K, by which $H_{\mbox{c2}}$ was calculated to be approximately 60 T if Werthamer-Helfand-Hohenberg (WHH) model is applied\cite{SakuraiHc2}. This value of the initial slope is close to that estimated from the specific heat data\cite{Cp}. However, by transport measurements, much smaller $H_{\mbox{c2}}$ has been obtained even for the case that the magnetic field was applied perpendicular to $c$-axis\cite{Sasaki}. In the present study, we prepared samples carefully and elucidated influence of synthesis conditions on the superconducting properties including $H_{\mbox{c2}}$.

\section{Experimental}
The powder samples of Na$_{x}$CoO$_{2}\cdot y$H$_{2}$O were synthesized from Na$_{0.7}$CoO$_{2}$ essentially in the same way as reported in the previous paper\cite{Nature}. Synthesis of the precursor of Na$_{0.7}$CoO$_{2}$ was described elsewhere\cite{SakuraiJPSJ}. The duration times of the immersion in Br$_{2}$/CH$_{3}$CN and water were both 5 days. Volume of water, $V_{\mbox{w}}$, was varied to find an optimum condition; each sample of Na$_{x}$CoO$_{2}$ ($x\sim 0.4$)\cite{TakadaJMC}, which was made from Na$_{0.7}$CoO$_{2}$ with the mass of 1 $g$, was immersed in water with $V_{\mbox{w}}=10$, 50, 100, 300, 500, or 1000 $ml$. The samples were filtered and stored in atmosphere of relative humidity of 70\%. The samples were characterized by powder X-ray diffraction (XRD) and inductive-coupled plasma atomic emission spectroscopy (ICP-AES). XRD measurements were carried out using a Bragg-Brentano-type diffractometer (RINT2200HF, Rigaku) with Cu $K_{\alpha}$ radiation. ICP-AES measurements were done by dissolving a sample in hydrochloric acid to determine the ratio of Na to Co, $x$. The magnetization, $M$, of the sample was measured using a commercial magnetometer with a superconducting quantum interference device (MPMS-XL, Quantum Design). Before each measurement under zero-field cooling (ZFC) condition, the magnetic field, $H$, was reset to 0 Oe at $T=10$ K or 300 K.

\section{Results and Discussion}
\begin{figure}[!t]
\begin{center}
\includegraphics[width=7cm,keepaspectratio]{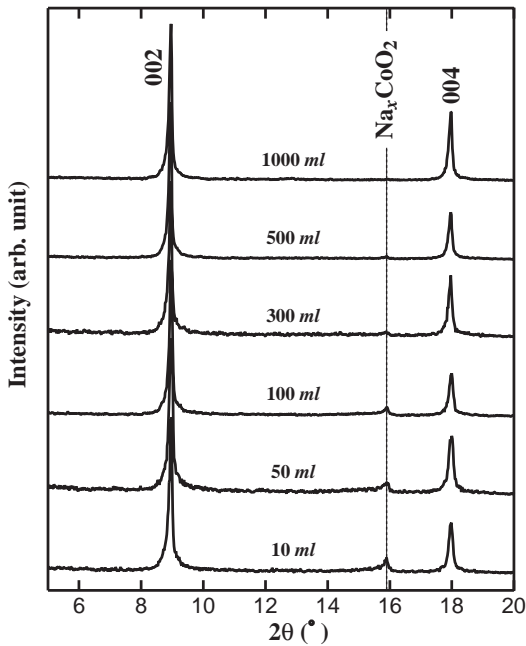}
\end{center}
\caption{
(a) XRD patterns of the samples immersed in water with $V_{\mbox{w}}$. The peak at $2\theta = 15.9$\degangle\ is assignable to Na$_{x}$CoO$_{2}$.
}
\label{XRD}
\end{figure}

\begin{figure}[!t]
\begin{center}
\includegraphics[width=7cm,keepaspectratio]{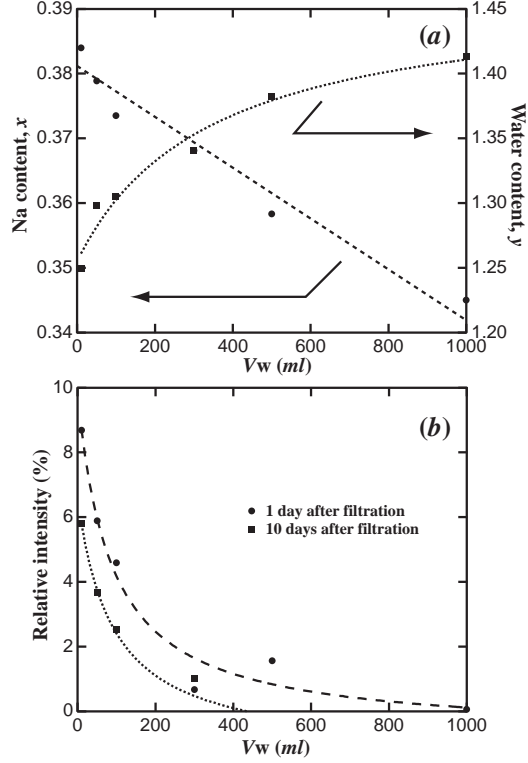}
\end{center}
\caption{
(a) The Na and water contents of the samples immersed in water with various $V_{\mbox{w}}$. The analyses was performed 1 day after the filtration. The broken and dotted lines are visual guides for $x$ and $y$, respectively. (b) Relative intensities of the peak at $2\theta = 15.9$\degangle\ caused by Na$_{x}$CoO$_{2}$ to the 002 peak of Na$_{x}$CoO$_{2}\cdot y$H$_{2}$O measured 1 day and 10 days after filtration. The broken and dotted lines are visual guides.
}
\label{Para}
\end{figure}

The XRD patterns of all the samples, which were measured just 1 day after the filtration, are shown in Fig. \ref{XRD}. As seen in the figure, all the samples immersed in water with $V_{\mbox{w}}\leq 500$ $ml$ included Na$_{x}$CoO$_{2}$ as an impurity phase. The Na content decreased with increasing $V_{\mbox{w}}$ as seen in Fig. \ref{Para}(a) which means that the Na ions were partly deintercalated in the hydration process\cite{TakadaJMC}. The pH of the water was nearly independent of $V_{\mbox{w}}$ being approximately 11 and this fact is consistent with the almost linear decrease of $x$ with increasing $V_{\mbox{w}}$. The hydration proceeded completely only for $x\leq ~0.35$, and thus for $V_{\mbox{w}} > 500$ $ml$, 

The relative intensity of the impurity peak at $2\theta =15.9$\degangle\ to the 002 peak of the hydrated phase is shown in Fig. \ref{Para}(b). After the sample was stored for 10 days under 70\% humidity, amount of the impurity of Na$_{x}$CoO$_{2}$ decreased in every sample and the 500 $ml$ sample became single phase. This suggests that the intercalation and/or deintercalation of water proceeded through the gas phase during the storage. Moreover, the Na atoms can also be deintercalated forming NaOH on the surface of the grains.  The 300 $ml$ sample did not become single phase even after it was stored for a month.

\begin{figure}[!t]
\begin{center}
\includegraphics[width=7cm,keepaspectratio]{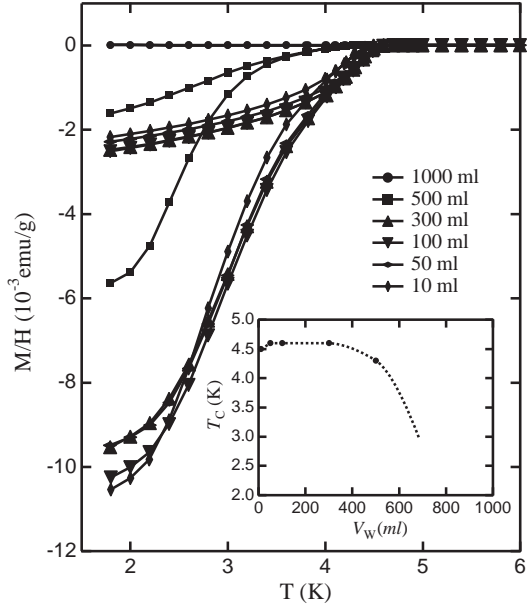}
\end{center}
\caption{
$M/H$-$T$ curves of the samples immersed in water with various $V_{\mbox{w}}$ measured a few days after filtration. The inset shows $T_{\mbox{c}}$ of the samples.
}
\label{chi}
\end{figure}

The magnetic susceptibility of each sample, which was measured within a few days after filtration, is shown in Fig. \ref{chi}. As seen in this figure, all the samples except the 1000 $ml$ one showed superconductivity. The variation of $T_{\mbox{c}}$ is shown as a function of $V_{\mbox{w}}$ in the inset. Obviously, the samples immersed in the water with $V_{\mbox{w}}\leq 300$ $ml$ showed better superconducting properties, although they are not single phase. On the other hand, the superconducting properties changed after the storage as seen in Fig. \ref{chiAft}. The superconducting properties of the 500 $ml$ and 1000 $ml$ samples were improved drastically after they were stored for 2 weeks, while those of the 300 $ml$ sample became rather bad after the 2 month storage. These facts are consistent with the idea that, during the storage, the intercalation and/or deintercalation of water can proceed through the gas and the Na atoms can also be deintercalated being extracted as NaOH. $T_{\mbox{c}}$ seems to depend on both the Na and the water content and there seems to be optimum values in them.  According to the present study, the best synthesis condition is the immersion to 500 $ml$ water for 1 $g$ of the starting oxide followed by 2 weeks storage under 70\% humidity.

\begin{figure}[!t]
\caption{
Variations of $M/H$-$T$ curves after the storage for the samples with $V_{\mbox{w}}=500$ $ml$ (a), 1000 $ml$ (b), and 300 $ml$ (c). 
}
\label{chiAft}
\end{figure}

The $M/H$-$T$ curves measured under various magnetic fields are shown in Fig. \ref{Hc2}. The samples A and B are different from those mentioned above. Each sample was single phases. The transition temperatures of samples A and B without magnetic field, $T_{\mbox{c}}(0)$, are approximately 4.2 and 4.6 K, respectively. As shown in Fig. \ref{Hc2}, $T_{\mbox{c}}$ under $H\geq 5$ T is hard to determine for the both samples, because the superconducting transitions are broad \cite{SakuraiHc2}. However, it is notable that the superconductivity of sample A is suppressed almost completely under 7 T while $M/H$ of sample B shows clear downturn below, at least, 3.5 K even under 7 T. Two explanations are possible for this result. First, $H_{\mbox{c2}}$ depends strongly on the sample quality and is improved greatly with slight increase of $T_{\mbox{c}}$. Second, since 7 T $M/H$-$T$ curves of samples A and B both start to deviate from the linear lines near $T_{\mbox{c}}(0)$ (see Fig. \ref{Hc2}), their $T_{\mbox{c}}$ do not decrease significantly under 7 T  but their superconducting volume fractions are different under 7 T. In either case, $H_{\mbox{c2}}$ seems to be quite high if it is measured for a high quality of sample. 
\begin{figure}[!t]
\caption{
$M/H$-$T$ curves of the samples A (a) and B (b) measured under various fields.  All the curves except those under 10 kOe are off-set to be distinguished. The dotted lines are visual guides.
}
\label{Hc2}
\end{figure}

The important thing is that the superconductivity under high magnetic field depends strongly on $T_{\mbox{c}}$, and $T_{\mbox{c}}$ is influenced by the synthesis conditions such as the volume of water in the hydration process or the storage time as mentioned above. This seems to be the reason for the inconsistent experimental results reported for the present system especially in the estimations of $H_{\mbox{c2}}$ and the NMR measurements. Our high quality samples with high $T_{\mbox{c}}$ values have been investigated by $\mu$SR, NMR, specific heat measurements, and  magnetic measurements\cite{Higemoto,Waki,SakuraiHc2,MichiokaK} to give consistent results. In $1/^{59}T_{1}T$ determined by NMR, superconducting transition is detected clearly with $T_{\mbox{c}}$ at approximately 4 K even under a high magnetic field of $\sim$7 T\cite{Kato} consistent with the high $H_{\mbox{c2}}$ estimated from the specific heat and magnetic measurements.

\section{Summary}
We synthesized powder samples of Na$_{x}$CoO$_{2}\cdot y$H$_{2}$O by changing the volume of water in the hydration process, and investigated their superconducting properties. The single phase of Na$_{x}$CoO$_{2}\cdot y$H$_{2}$O was obtained when the volume of water was 500 $ml$ and 1000 $ml$ per 1 g of starting materials of Na$_{0.7}$CoO$_{2}$. $T_{\mbox{c}}$ of the sample changed gradually while it was stored in the atmosphere of the humidity of 70\%. The superconducting properties under high magnetic field were very sensitive to $T_{\mbox{c}}$, and thus, to the synthesis conditions. These facts suggest that discrepancies in experimental results for the present compound result from the difference in sample quality.  $H_{\mbox{c2}}$ is obviously higher than 7 T in a high quality sample.

\section*{Acknowledge}
Special thanks to S. Takenouchi (NIMS) for his chemical analyses. We would like to thank H. D. Yang (National Sun Yat Sen University), J.-Y. Lin (National Chiao Tung University), F. Izumi, R. A. Dilanian, A. Tanaka (NIMS), K. Yoshimura, M. Kato, C. Michioka, T. Waki, K. Ishida (Kyoto University), W. Higemoto (JAERI), for useful discussion. This study was partially supported by Grants-in-Aid for Scientific Research (B) from Japan Society for the Promotion of Science (16340111). One of the authors (H.S) is research fellow of the Japan Society for the Promotion of Science.

\end{document}